\documentclass[fleqn,twoside]{article}
\usepackage{amssymb}
\usepackage{macros}
\usepackage{graphicx}
\usepackage{espcrc2}

\title{Universality and RG-improved gauge actions}
\author{Silvia Necco\address{DESY, Platanenallee 6, D-15738 Zeuthen, 
Germany}\thanks{Supported by the European Community's Human potential programme under HPRN-CT-2000-00145 Hadrons/LatticeQCD. We thank NIC (J\"ulich) for granting CPU-resources for this project.}}
\begin{document}
\begin{abstract}
The reference energy scale $r_{0}$ is evaluated for RG-improved Iwasaki and DBW2 gauge actions ($N_{f}=0$), at values of the deconfinement coupling $\beta_{c}$ corresponding to $N_{t}=3,4,6,8$. The universality of $r_{0}T_{c}$ between Iwasaki and Wilson action is confirmed; the scaling behaviour and the influence of the violation of positivity in the extraction of effective masses are investigated.
\end{abstract}
\maketitle
\vspace{-0.5cm}
\vspace{-0.5cm}
\section{INTRODUCTION}
In the last years the lattice comunity made a big effort in the improvement of
both fermionic and gauge actions. Concerning the latter, the tree-level and
one-loop improved action were introduced in the Symanzik framework \cite{Luscher:1985zq}.  
Iwasaki \cite{Iwasaki:1983ck} and the QCD-TAR0 collaboration \cite{Takaishi:1996xj} followed another approach based on the renormalization group (RG); the improvement is obtained estimating the renormalized trajectory, in \cite{Iwasaki:1983ck} through perturbation theory, in \cite{Takaishi:1996xj} using a Schwinger-Dyson method.\\
The Iwasaki action has been used by the CP-PACS collaboration \cite{AliKhan:2001tx} for the most advanced computation of the light hadron spectrum; moreover,
the RG-improved actions have been considered as candidates to be used
in the next simulations on Ginsparg-Wilson/domain wall fermions, due
to the suppression of small instantons and dislocations and a possible
remedy of the problem of residual chiral symmetry breaking for
domain wall fermions \cite{Orginos:2001xa,Aoki:2001dx}.\\
This motivates more investigations about their properties,
in particular the universality between the Iwasaki and the
standard plaquette action, the scaling properties and the potential
problem in the evaluation of the physical observables due to the
violation of physical positivity.
\vspace{-0.5cm}

\section{IMPROVED GAUGE ACTIONS AND LOSS OF PHYSICAL POSITIVITY}
\begin{figure}
\includegraphics[width=6cm]{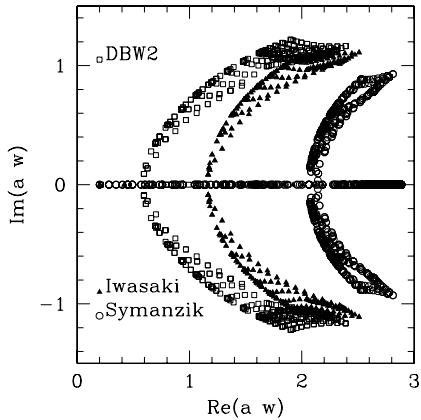}
\vspace{-1cm}
\caption{The poles of the propagator for different actions}\label{poli}
\vspace{-0.5cm}
\end{figure}

\begin{figure}
\includegraphics[width=6cm]{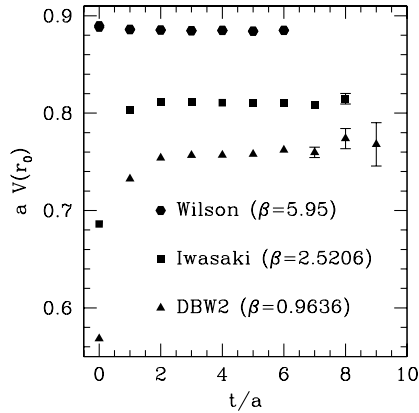}
\vspace{-1.5cm}
\caption{The effective potential at $r\sim r_{0}$ for different actions, from smeared Wilson loops at smearing level $\sim (r_{0}/{a})^{2}.$}\label{pot_r0}
\vspace{-0.5cm}
\end{figure}

We will consider improved gauge actions of the form
\begin{equation}
\Sg=\frac{\beta}{3}\big(\sum_{x;\mu<\nu}(1-8c_{1})P_{\mu\nu}^{1\times
      1}+c_{1}\sum_{x;\mu\neq\nu}P_{\mu\nu}^{1\times 2}\big),
\end{equation}
where $P_{\mu\nu}^{1\times 1}$ is the usual plaquette term and
$P_{\mu\nu}^{1\times 2}$ contains all loops enclosing two coplanar
plaquettes. The coefficient $c_{1}$ takes different values for
various choices of the improved actions
$$
c_{1} =\left\{\begin{array}{ll}
-1/12  & \textrm{Symanzik, tree level impr.}\\
-0.331 & \textrm{Iwasaki, RG}\\
-1.4088 & \textrm{DBW2 (QCD-TAR0), RG}
\end{array}\right.
$$
In general, adding operators of dimension 6 to the standard
Wilson action violates the physical positivity at finite lattice
spacing. \\
A transfer matrix, $\trans$, can be nevertheless defined, but
complex eigenvalues  may occur and there are
contributions in the spectral decomposition of
two-point functions with negative weight \cite{Luscher:1984is}.
Near the continuum limit, positivity is only lost at energies of the
order of the cutoff: if we
restrict ourselves to the states with
small energy relative to the cutoff the physical positivity is completely restored.
Perturbative and free field calculations suggest there exists a
$0<\epsilon<1$ such that independently of the cutoff all spectral values $\lambda$ of $\trans$ with
$|\lambda|\geq\epsilon\Lambda$ are real and positive, where $\Lambda$
is the largest eigenvalue of $\trans$. These eigenvalues $\lambda$ can be interpreted as energy values through 
\begin{equation}
E=-\frac{1}{a}\ln(\lambda/\Lambda).
\end{equation}
An estimate of $\epsilon$ can be obtained in weak coupling
perturbation theory by determining the location of unphysical (but
gauge independent) poles in the propagator.
The general solution will be
$w(k_{1},k_{2},k_{3})=\mathfrak{Re}(w)+i\mathfrak{Im}(w)$,
with $(k_{1},k_{2},k_{3}) \in $ Brillouin zone. In \fig{poli} the
solutions for different improved actions are shown; 
for $\mathfrak{Re}(w) < E_{max}$ one has $\mathfrak{Im}(w)=0$ for all
solutions at a given momentum. The value $aE_{max}=-\ln(\epsilon)$ is related to the
parameter $c_{1}$ by the relation:
\begin{equation}
a E_{max} =2\textrm{arcsinh} \left(\frac{1}{2\sqrt{-2 c_{1} }}\right).
\end{equation}
One estimates that the contribution of the
unphysical states in asymptotic decay of euclidean 2-point functions
can then be neglected if
\begin{equation}
t/a\gg t_{min}/a  =\frac{1}{aE_{max}}=\left\{\begin{array}{ll}
0.485 & \textrm{Symanzik}\\
0.860 & \textrm{Iwasaki}\\
1.703 & \textrm{DBW2}
\end{array}\right.
\end{equation}
This turns
out to play a r\^ole also in MC simulations; \fig{pot_r0} shows the
effective potential at a fixed distance $r\sim r_{0}$ for the
different actions. The APE smearing procedure \cite{Albanese:ds} was used, and the smearing level considered in the plot is the one estimated to be optimal for the Wilson plaquette action \cite{Guagnelli:1998ud}.
Negative contributions at small $t/a$ are strong and they
can be a problem for the applicability of the variational
method \cite{Campbell:1987nv} to extract the mass spectrum. 
That procedure is well defined only for positive definite correlation functions, but this is verified only if $t\gg t_{min}$, where the statistical fluctuations can be large. 
Moreover, although one can observe a plateau in  \fig{pot_r0}, it is clear that for the improved actions there is no unambiguous criterion for the choice of the smearing operators.

\section{SCALING OF $T_{c}r_{0}$}
\begin{figure}
\includegraphics[width=7cm]{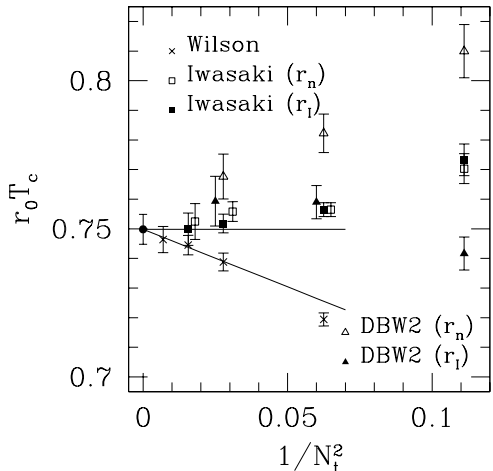}
\vspace{-1.3cm}
\caption{$T_{c}r_{0}$ for different actions. The $x$ coordinates were slighly shifted for clarity reasons. }\label{tcr0}
\vspace{-0.7cm}
\end{figure}

The deconfinement transition occurs at the critical temperature
\be
T_{c}=\frac{1}{N_{t}a(\beta_{c})}.
\ee
For known values of $\beta_{c}$ at given $N_{t}$ for different actions one can refer to \cite{Boyd:1996bx,Beinlich:1997ia,Okamoto:1999hi,deForcrand:1999bi}.
Expressing the critical temperature in units of the string tension,
the following results were obtained
\begin{equation}
 T_{c}/\sqrt{\sigma} =\left\{\begin{array}{ll}
 0.630(5)\quad  (N_{t}=\infty) & \textrm{Wilson}\cite{Beinlich:1997ia}\\
 0.650(5)\quad  (N_{t}=\infty) & \textrm{Iwasaki}\cite{Okamoto:1999hi}\\
 0.6301(65)\quad (N_{t}=6) & \textrm{DBW2}\cite{deForcrand:1999bi}
\end{array}\right.
\end{equation}
For the Iwasaki and the Wilson action a difference of the order
4$\sigma$ in the continuum results is observed. It is unrealistic that
the origin of the discrepancy is a violation of the universality; the
most natural explanation is that the string tension is difficult to
determine, above all at small lattice spacings. We decided to evaluate instead the Sommer scale $r_{0}$ \cite{Sommer:1993ce} defined through the force between a pair of static quarks, $F(r_{0})r_{0}^2=1.65$, at $\beta_{c}$.
For the Wilson action a parametrization formula for $5.7\leq\beta\leq 6.57$
 \cite{Guagnelli:1998ud} which relates $r_{0}/a$ to $\beta$ can be used;
for the Iwasaki and DBW2 action there is no available evaluation of $r_{0}$, so it was measured from Wilson loop correlation matrices, using 
\begin{equation}
\frac{V(r)-V(r-a)}{a}  =\left\{\begin{array}{ll}
 F\left(r-\frac{a}{2}\right) & \textrm{naive definition}\\
 F(r_{I})  & \textrm{tree-level improved},
\end{array}\right.
\end{equation}
where $r_{I}$ is defined through the scalar lattice propagator in 3 dimensions,
$$
(4\pi  r_{I}  ^{2})^{-1}=[G(r,0,0)-G(r-a,0,0)]/a.
$$
The operator corresponding to the best smearing for the Wilson action was considered; due to the negative contributions in the correlation functions and the previous discussion, we decided to avoid the variational method.\\
The results for $T_{c}r_{0}$, together with the continuum extrapolation, are shown in \fig{tcr0}.
For the Iwasaki action there is
no appreciable difference between the results obtained with
$r_{naive}$ and $r_{I}$; the data show better scaling properties in
comparison to the Wilson action and moreover the universality is confirmed.
Also for the $DBW2$ action the scaling properties are improved,
although only using $r_{I}$ instead of $r_{naive}$.
A constrained fit 
\begin{equation}
T_{c}r_{0}=T_{c}r_{0}|_{a=0}+s\cdot\left(\frac{a}{T_{c}}\right)^{2}, \textrm{with}\quad N_t\geq 6
\end{equation}
for Iwasaki and Wilson action yields the continuum result
\begin{equation}
T_{c}r_{0}=0.7498(50).
\end{equation} 
At $N_{t}=6$ the Wilson action shows scaling violations for $r_{0}T_{c}$ of 
about $1.5\%$, while they are $0.3\%$ for the Iwasaki action.

\section{CONCLUSIONS}
The loss of physical positivity can be a serious problem for the
extraction of the physical observables, 
in particular the variational technique is not easily applicable. For the
Iwasaki and the Wilson action the universality of $T_{c}r_{0}$ is demonstrated. In general the lattice artefacts seem to be reduced for the Iwasaki
action; it would be anyhow useful to have more investigations on the
scaling behavior, for example evaluating glueball masses.

\end{document}